\documentclass[twocolumn,showpacs,preprintnumbers,amsmath,amssymb]{revtex4}

\usepackage{graphicx}
\usepackage{dcolumn}
\usepackage{bm}

\def\gsim{\mathop {\vtop {\ialign {##\crcr 
$\hfil \displaystyle {>}\hfil $\crcr \noalign {\kern1pt \nointerlineskip } 
$\,\sim$ \crcr \noalign {\kern1pt}}}}\limits}
\def\lsim{\mathop {\vtop {\ialign {##\crcr 
$\hfil \displaystyle {<}\hfil $\crcr \noalign {\kern1pt \nointerlineskip } 
$\,\,\sim$ \crcr \noalign {\kern1pt}}}}\limits}

\begin{document}

\preprint{APS/123-QED}

\title{Enhancement of Nuclear Spin-Lattice Relaxation Rate and Spin Susceptibility 
\\
due to Valence Fluctuations 
\\
-Origin of Anomalously Enhanced Wilson Ratio in Ce and Yb Systems-
}

\author{Shinji Watanabe$^1$}
\author{Kazumasa Miyake$^2$}%
\affiliation{%
Department of Applied Physics, University of Tokyo, Hongo 7-3-1, Bunkyo-ku, Tokyo, 113-8656, Japan$^1$
\\
Division of Materials Physics, Department of Materials Engineering Science, Graduate School of
Engineering Science, Osaka University, Toyonaka, Osaka 560-8531, Japan$^2$
}%

\date{November 24, 2008}

\begin{abstract}
We show theoretically that the nuclear spin-lattice relaxation rate $(T_1T)^{-1}$ and 
the spin susceptibility $\chi_{\rm s}(T)$ exhibit divergent behaviors toward zero temperature 
at the quantum critical point (QCP) of the first-order valence transition. 
Remarkable enhancement in $(T_1T)^{-1}$ and $\chi_{\rm s}(T)$ 
is induced by valence fluctuations 
even at the valence-crossover temperature far away from the QCP. 
This mechanism well explains peculiar behaviors observed recently 
in $\rm YbAuCu_4$ and also gives a systematic explanation for 
$\rm YbXCu_4$ for X=In, Au, Ag, Tl, and Pd from the viewpoint of the closeness to the QCP. 
This also explains anomalously enhanced Wilson ratio observed 
in the paramagnetic Ce and Yb based compounds. 
This offers a new concept that spin fluctuations are induced via relative charge fluctuations, 
which can be generally applied to the systems with valence instabilities.

\end{abstract}

\maketitle
 
Quantum critical phenomena in strongly-correlated electron systems 
have been discussed extensively in the context of magnetic phase 
transitions ~\cite{Moriya,Hertz,Millis}. 
Recently, instabilities in charge sectors have attracted 
much attention, since underlying influence of valence instability 
is suggested by variety of materials~\cite{M07,Yuan,Holmes,WIM}: 
Importance of critical valence fluctuations has been argued 
as a possible origin of anomalies such as 
$T$-linear resistivity, enhanced residual resistivity 
and superconductivity for the materials with valence-fluctuating ions 
such as Ce and Yb.

Valence transition is isostructural transition 
as known as $\gamma$-$\alpha$ transition in Ce metal~\cite{Cevalence} 
and also in $\rm YbInCu_4$~\cite{Felner} characterized by a jump of 
the valence of the Ce and Yb ion.  
In $\rm YbInCu_4$ the first-order valence transition at $T=42$~K takes place 
with the valence of Yb being +2.97 (+2.84) in the high (low)-temperature phase~\cite{YbInCu4v}. 
Interestingly, 
anomalous behaviors in the $\rm YbXCu_4$ have been revealed by 
systematic measurements: 
A remarkable enhancement in the nuclear spin-lattice relaxation rate $(T_{1}T)^{-1}$ 
has been discovered recently for X=Au~\cite{wada,wada2}. 
A mysterious behavior of this material has been recognized in the spin susceptibility, 
which increases enormously toward zero temperature~\cite{Sarrao} 
in spite of the antiferromagnetically-ordered ground state 
with $T_{\rm N}\sim0.8$~K~\cite{Bauer}. 
By applying the magnetic field $T_{\rm N}$ is suppressed to 0~K at $H_{\rm c}\sim 1.3$~T 
and a remarkable point is that by further applying $H$ 
the enhancement of $(T_1T)^{-1}$ emerges at finite temperature, $T=T_{\rm v}(H)$~\cite{wada,wada2}. 
Since these anomalies appear even far away from $H_{\rm c}$, i.e., 
in the regime where the magnetic order is completely destroyed, 
this is not due to the magnetic fluctuations. 

Furthermore, it has been discovered that the $\rm ^{63}Cu$ NQR frequency $\nu_{\rm Q}$ 
shows a sharp drop at $T_{\rm v}$ when $T$ decreases~\cite{wada2}. 
Since $\nu_{\rm Q}$ measures the charge distribution of the Yb and surrounding ions, 
the change of $\nu_{\rm Q}$ indicates that the Yb valence changes at $T_{\rm v}$ sharply. 
A peak structure in the spin susceptibility $\chi_{\rm s}(T)$ at $T_{\rm v}$ 
has been also detected for X=Ag~\cite{Sarrao} and Tl~\cite{Sarrao}, 
and enhancement in $(T_1T)^{-1}$ has been also found at $T_{\rm v}$ 
for X=Ag~\cite{Otani} and X=Pd~\cite{YbPdCu4} recently. 
These observations suggest that this behavior is not specific to a special material, 
but is rather universal. 
Since valence fluctuations are ascribed to the relative charge fluctuations 
between f and conduction electrons, these observations challenge the conventional concept that 
the nuclear spin-lattice relaxation rate and the spin susceptibility reflect magnetic fluctuations 
in the paramagnetic-metal phase. 

In this Letter, we resolve this puzzle by showing that 
valence fluctuations indeed induce spin fluctuations. 
We show that 
$(T_1T)^{-1}$ as well as $\chi_{\rm s}(T)$ shows divergence  toward zero temperature 
at the quantum critical point (QCP) of the valence transition. 
Even in the valence-crossover region at finite temperatures away from the QCP, 
$(T_1T)^{-1}$ and $\chi_{\rm s}(T)$ are shown to be enhanced. 
This mechanism gives a systematic explanation for peculiar behaviors observed in $\rm YbXCu_4$ 
for X=In, Au, Ag, Tl and Pd, and also accounts for anomalously enhanced Wilson ratio 
observed in several Yb and Ce materials located near 
the QCP of the valence transition. 
This result offers a new concept that spin fluctuations are induced via 
relative charge fluctuations, which can be generally applied to the systems with valence instabilities. 

Let us start our analysis by introducing a minimal model 
which describes the essential part of the Yb and Ce systems 
in the standard notation~\cite{Falikov,Varma}: 
\begin{equation}
H=H_{\rm c}+H_{\rm f}+H_{\rm hyb}+H_{U_{\rm fc}}, 
\label{eq:PAM} 
\end{equation}
where 
$H_{\rm c}=\sum_{{\bf k}\sigma}\varepsilon_{\bf k}
c_{{\bf k}\sigma}^{\dagger}c_{{\bf k}\sigma}$, 
$H_{\rm f}=\varepsilon_{ \rm f}\sum_{i\sigma}n^{ \rm f}_{i\sigma}
+U_{\rm ff}\sum_{i=1}^{N}n_{i\uparrow}^{ \rm f}n_{i\downarrow}^{ \rm f}
$, 
$H_{\rm hyb}=\sum_{{\bf k}\sigma}V_{\bf k}\left(
f_{{\bf k}\sigma}^{\dagger}c_{{\bf k}\sigma}+c_{{\bf k}\sigma}^{\dagger}f_{{\bf k}\sigma}
\right)$ 
and 
$
H_{U_{\rm fc}}=
U_{\rm fc}\sum_{i=1}^{N}n_{i}^{ \rm f}n_{i}^{ c}
$. 
The $U_{\rm fc}$ term is the Coulomb repulsion between f and conduction electrons, 
which is considered to play an important role in the valence transition: 
In the case of Ce metal which exhibits the $\gamma$-$\alpha$ transition, 
the 4f- and 5d-electron bands are located at the Fermi level~\cite{Ceband}.   
Since both the orbitals are located on the same Ce site, this term cannot be neglected. 
For $\rm YbInCu_4$, the considerable magnitude of the In 5p and Yb 4f hybridization 
was pointed out by the band-structure calculation~\cite{Takegahara} 
and recent high-resolution photoemission spectra has detected a remarkable 
increase of the p-f hybridization at the first-order valence transition~\cite{Yoshikawa}.
These results suggest importance of $V_{\bf k}$ and $U_{\rm fc}$. 
Actually, the reason why the critical-end temperature is so high as much as 600~K 
in Ce metal in contrast to that in $\rm YbInCu_4$ can be understood in terms 
of $U_{\rm fc}$: 
In $\rm YbInCu_4$, $U_{\rm fc}$ is the intersite interaction, which should be 
smaller than that of Ce metal. 
This view also gives an explanation for the reason why most of Ce and Yb compounds 
only shows the valence crossover. 
Namely, most of the compounds seems to have a moderate value of $U_{\rm fc}$ 
due to its intersite origin, which is smaller than the critical value 
to cause a jump of the valence. 
However, even in the valence-crossover regime, 
underlying influence of the valence instability 
causes intriguing phenomena as shown below. 
It is noted that importance of $U_{\rm fc}$ has been discussed 
by several authors 
for $\rm YbInCu_4$~\cite{Zlatic,Ufc}. 

\begin{figure}
\includegraphics[width=67mm]{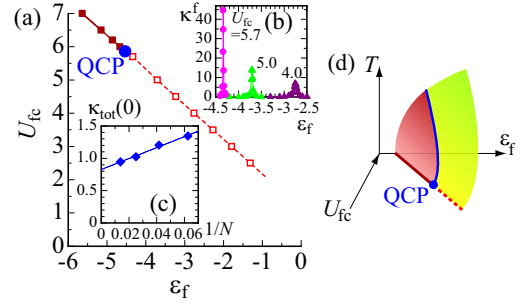}
\caption{\label{fig:PD} 
(Color) 
(a) Ground-state phase diagram determined by the DMRG 
for $\varepsilon_{\bf k}=-2\cos(k)$, $V=V_{\bf k}=0.1$ and $U=100$ 
at filling $\sum_{i=1}^{N}\langle n^{\rm f}_i+n^{\rm c}_i\rangle/(2N)=7/8$~\cite{WIM}. 
(b) Valence susceptibility has a peak on the dashed line in (a). 
(c) System-size dependence of the total-charge compressibility at the QCP 
$(\varepsilon_{\rm f}, U_{\rm fc})=(-4.5206, 5.8460)$.
(d) Schematic phase diagram in the $\varepsilon_{\rm f}$-$U_{\rm fc}$-$T$ space 
for a certain $V_{\bf k}$ and $U_{\rm ff}(>U_{\rm fc})$. 
The first-order valence transition surface (brown surface) 
with the critical end line (blue line) touched on $T=0$ at the 
QCP continues to the valence-crossover surface 
(light-green surface). 
The valence-crossover surface at which $\kappa^{\rm f}(T)$ has a maximum 
is denoted as $T_{\rm v}$ (see text). 
}
\end{figure}

In this model (\ref{eq:PAM}) 
the first-order transition between the larger $\langle n_{\rm f}\rangle$ 
and the smaller $\langle n_{\rm f}\rangle$ is caused by $U_{\rm fc}$, 
since the large $U_{\rm fc}$ forces electrons to pour into either the f level or 
the conduction band~\cite{OM,WIM}. 
Figure~\ref{fig:PD}(a) shows the ground-state phase diagram 
in the $\varepsilon_{\rm f}$-$U_{\rm fc}$ plane 
determined by the density-matrix renormalization group (DMRG) 
for $\varepsilon_{\bf k}=-2\cos(k)$, 
$V=V_{\bf k}=0.1$ and $U=100$ 
at filling $\sum_{i=1}^{N}\langle n^{\rm f}_i+n^{\rm c}_i\rangle/(2N)=7/8$~\cite{WIM}. 
We note that 
essentially the same phase diagram has been obtained in the infinite dimensional system~\cite{saiga}. 
The first-order-transition line (brown line) separates 
the paramagnetic-metal phase, namely, 
the larger $\langle n_{\rm f}\rangle$ phase in the smaller $\varepsilon_{\rm f}$ 
and $U_{\rm fc}$ regime, and the smaller $\langle n_{\rm f}\rangle$ phase 
in the larger $\varepsilon_{\rm f}$ and $U_{\rm fc}$ regime. 
As $U_{\rm fc}$ decreases the jump in $n_{\rm f}$ at the first-order 
transition decreases and terminates at the QCP, 
at which the valence fluctuation diverges. 

\begin{figure}
\includegraphics[width=73mm]{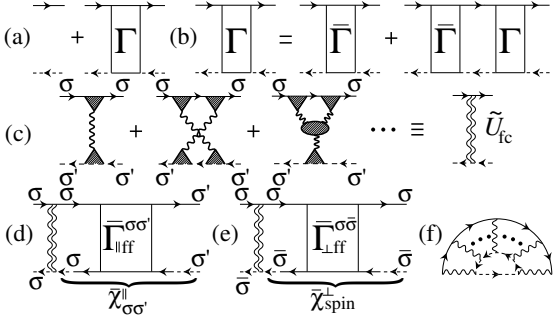} 
\caption{\label{fig:RPA}
(a) The most dominant part near the QCP of $\chi^{\rm ff}$ and $\chi^{\rm ff}_{+-}$ 
with four-point vertex $\tilde{\Gamma}$ and (b) its integral equation. 
(c) Vertex parts including $U_{\rm ff}$ (hatched area) connected by $U_{\rm fc}$ (wiggly line). 
All contribution of this type is represented by the double wiggly line as $\tilde{U}_{\rm fc}$. 
(d) $\tilde{\Gamma}$ for $\chi^{\rm ff}$ and (e) for $\chi^{\rm ff}_{+-}$. 
$\bar{\sigma}$ is antiparallel spin to $\sigma$. 
$\bar{\Gamma}_{\rm ff}$ contains all contributions other than the diagrams expressed in (c). 
(f) RPA-type self energy for $U_{\rm fc}$. 
The solid line and dashed line represent 
$G^{\rm ff}_{\sigma}$ and $G^{\rm cc}_{\sigma}$, respectively. 
}
\end{figure}

The valence fluctuation is measured by 
the dynamical valence susceptibility defined by 
%
$
\chi^{\rm ff}({\bf q},i\omega_n)
\equiv\int_{0}^{\beta}d\tau\langle 
T_{\tau}
n^{\rm f}({\bf q},\tau)n^{\rm f}({\bf -q},0)\rangle
{\rm e}^{i\omega_n\tau} 
$
%
with $\omega_n=2n\pi T$ $(n=0,\pm1,\pm2,\cdots)$. 
The most dominant part of $\chi^{\rm ff}$ near the QCP 
is expressed by the four-point vertex function $\Gamma$ in Fig.~\ref{fig:RPA}(a), 
which satisfies the integral equation shown in Fig.~\ref{fig:RPA}(b). 
Here, the solid line and dashed line represent quasiparticle parts of Green functions 
near the Fermi level 
$G^{\rm ff}_{\sigma}({\bf q},i\varepsilon_{n})\sim 
1/(i\varepsilon_{n} -E_{{\bf q}\sigma}^{*})$, 
$G^{\rm cc}_{\sigma}({\bf q},i\varepsilon_{n})\sim 
1/(i\varepsilon_{n} -E_{{\bf q}\sigma}^{*})$, 
respectively, 
where $\varepsilon_{n}=(2n+1)\pi T$ and 
the origin of the energy is set as $\mu\equiv 0$. 
Here, 
$E_{{\bf q}\sigma}^{*}$ satisfies 
$
(z-\varepsilon_{\rm f}-\Sigma^{\rm ff}_{{\bf q}\sigma}(z))
(z-\varepsilon_{\bf q})-
|\tilde{V}_{{\bf q}\sigma}(z)|^2
=0 
$ 
$(z=E_{{\bf q}\sigma}^{*})$, 
where 
$\tilde{V}_{{\bf q}\sigma}(z)\equiv 
V_{\bf q}+\Sigma^{\rm fc}_{{\bf q}\sigma}(z)
$ 
with $\Sigma^{\rm ff}_{{\bf q}\sigma}$ and 
$\Sigma^{\rm fc}_{{\bf q}\sigma}$ being the 
self energies due to $U_{\rm ff}$ and $U_{\rm fc}$, respectively. 
The effect of 
the renormalization amplitude defined by
$
a_{{\bf q}\sigma}^{\rm ff}\equiv 
[
1-
\partial \Sigma^{\rm ff}_{{\bf q}\sigma}(\varepsilon)
/\partial\varepsilon
+|\tilde{V}_{{\bf q}\sigma}(\varepsilon)|^2
/(\varepsilon-\varepsilon_{\bf q})^{2}
-2{\rm Re}[
\tilde{V}^{*}_{{\bf q}\sigma}(\varepsilon)
\partial \Sigma^{\rm fc}_{{\bf q}\sigma}(\varepsilon)
/\partial\varepsilon
]
/(\varepsilon-\varepsilon_{\bf q})
]^{-1}|_{\varepsilon=0}
$ 
and 
$
a_{{\bf q}\sigma}^{\rm cc}\equiv 
a_{{\bf q}\sigma}^{\rm ff}
|\tilde{V}_{{\bf q}\sigma}(0)|^{2}
/\varepsilon_{\bf q}^{2}
$ 
is absorbed in the definition of vertex functions, internal and 
external~\cite{AGD}. 
The vertex $\bar{\Gamma}$ consists of 
the interactions via $U_{\rm fc}$ with vertex parts including $U_{\rm ff}$, 
whose contributions are represented by the double wigly line 
as $\tilde{U}_{\rm fc}$ in Fig.~\ref{fig:RPA}(c) 
and all the other contributions such as entangled $U_{\rm fc}$ and $U_{\rm ff}$ 
shown as $\bar{\Gamma}^{\sigma\sigma'}_{|| {\rm ff}}$ 
in Fig.~\ref{fig:RPA}(d). 
Here, $\bar{\Gamma}=\hat{a}\tilde{U}_{\rm fc}\hat{I}\hat{\chi}^{||}$, 
where 
$\hat{a}$ and 
$\hat{\chi}^{||}$ are the $2\times 2$ matrices with elements 
$[\hat{a}]_{\sigma\sigma'}\equiv a^{\rm ff}_{\sigma}a^{\rm cc}_{\sigma'}$ 
and 
$[\hat{\chi}^{||}]_{\sigma\sigma'}=\bar{\chi}_{\sigma\sigma'}$ 
defined as Fig~\ref{fig:RPA}(d), respectively. 
$\hat{I}$ is the unit matrix. 
Then, we have 
$\chi^{\rm ff}\propto
{\rm Tr}[\hat{\chi}^{||}(\hat{I}-\tilde{U}_{\rm fc}\hat{I}\hat{\chi}^{||})^{-1}]$. 
By using the relation 
$\bar{\chi}_{\rm charge}\equiv
\bar{\chi}_{\uparrow\uparrow}+\bar{\chi}_{\uparrow\downarrow}$ 
and 
$\bar{\chi}_{\rm spin}^{||}\equiv
\bar{\chi}_{\uparrow\uparrow}-\bar{\chi}_{\uparrow\downarrow}$ 
and noting the fact that $\bar{\chi}_{\rm charge}\ll\bar{\chi}^{||}_{\rm spin}$ 
holds for typical heavy-electron systems~\cite{Kyamada}, we have 
$\chi^{\rm ff}\propto
\bar{\chi}_{\rm spin}^{||}/(1-\tilde{U}_{\rm fc}\bar{\chi}_{\rm spin}^{||})$. 

Figure~\ref{fig:PD}(d) illustrates the schematic phase diagram 
in the $T$-$\varepsilon_{\rm f}$-$U_{\rm fc}$ space. 
At the critical end line of the first-order transition 
illustrated by the blue line in Fig.~\ref{fig:PD}(d) 
where the denominator of $\chi^{\rm ff}$ equals zero, 
$1-\tilde{U}_{\rm fc}\bar{\chi}_{\rm spin}^{||}({\bf 0},0)=0$, 
the valence susceptibility 
$\kappa^{\rm f}\equiv\chi^{\rm ff}({\bf 0},0)=
-\partial \langle n^{\rm f}\rangle/\partial \varepsilon_{\rm f}$ 
diverges. 
Note here that $\kappa^{\rm f}$ 
shows not only divergence at the critical-end line, 
but also has a maximum at the valence-crossover surface 
($\kappa^{\rm f}$ has a peak in Fig.~\ref{fig:PD}(b) 
on the dashed line in Fig.~\ref{fig:PD}(a)), 
which is illustrated as the surface extended from the first-order one 
in Fig.~\ref{fig:PD}(d)~\cite{OM,WIM}. 
This implies that even if the system is away from the critical end line, 
the valence fluctuation is enhanced at the crossover surface $T_{\rm v}$ 
in the parameter space of $T$ and pressure and/or chemical doping. 
Furthermore, it has been revealed recently that even in the valence-crossover 
regime, the critical point is induced by applying the magnetic field~\cite{WTMF}. 
This explains the role of the magnetic field in the $T$-$H$ phase diagram of 
$\rm YbAuCu_4$~\cite{wada} and $\rm YbPdCu_4$~\cite{YbPdCu4}. 
Namely, the magnetic field not only destroys the magnetic order, but also 
plays a role to make the valence-crossover temperature $T_{\rm v}$ finite. 
This is consistent with the experimental fact that $T_{\rm v}(H)$ emerges 
under $H$ even much larger than $H_{\rm c}$ in the $T$-$H$ plane, 
at which $\nu_{\rm Q}$ shows a sharp change.

Now let us focus on the nuclear spin-lattice relaxation rate 
defined by 
%
$
\frac{1}{T_1}=\frac{\gamma_{n}^2k_{\rm B}T}{{g_{\rm f}}^2\mu_{\rm B}^2}
\sum_{\bf q}|\tilde{A}_{\bf q}|^2\frac{{\rm Im}
\chi_{+-}^{\rm ff}({\bf q},i\omega_0)}{\omega_0},
$
%
where 
$\gamma_{\rm n}$ is the gyromagnetic ratio of nuclear spin, 
$g_{\rm f}$ is a Lande's g factor for f electrons and 
$\tilde{A}_{\bf q}$ is the hyperfine-coupling constant. 
The dynamical f-spin susceptibility is defined by
$
\chi_{+-}^{\rm ff}({\bf q},i\omega_n)
\equiv\int_{0}^{\beta}d\tau
\langle T_{\tau}S^{\rm f}_{+}
({\bf q},\tau)S^{\rm f}_{-}({\bf -q},0)\rangle
{\rm e}^{i\omega_n\tau}.
$
We note here that $\chi^{\rm ff}_{+-}$ has essentially the same structure 
as $\chi^{\rm ff}$, whose most dominant terms near the QCP 
is expressed in Figs.~\ref{fig:RPA}(a) and (b).  
The vertex $\bar{\Gamma}$ in Fig.~\ref{fig:RPA}(b) is given by 
$\bar{\Gamma}=a^{\rm ff}_{\sigma}a^{\rm cc}_{\bar{\sigma}}
\tilde{U}_{\rm fc}\bar{\chi}_{\rm spin}^{\perp}$ 
as shown in Figs.~\ref{fig:RPA}(e) and (c). 
Then, we have $\chi^{\rm ff}_{+-}\propto
\bar{\chi}_{\rm spin}^{\perp}/(1-\tilde{U}_{\rm fc}\bar{\chi}_{\rm spin}^{\perp})$. 
Since SU(2) symmetry of the system ensures 
$\bar{\chi}_{\rm spin}^{\perp}=\bar{\chi}_{\rm spin}^{||}$, 
it turns out that $\chi^{\rm ff}_{+-}$ has the same form as $\chi^{\rm ff}$. 
Hence, at the critical-end line as well as the QCP of 
the valence transition (blue line in Fig~\ref{fig:PD}(d)), 
$1-\tilde{U}_{\rm fc}\bar{\chi}^{\perp}_{\rm spin}({\bf 0},0)=0$ holds, 
which makes $\chi^{\rm ff}_{+-}({\bf 0},0)$ diverge~\cite{chipm}. 
Since the spectral weight of f electrons is dominated by the incoherent part 
around $\varepsilon\sim\varepsilon_{\rm f}$, 
the coherent part responsible for quasiparticles amounts to a tiny contribution 
in the order of $a^{\rm ff}_{{\bf k}_{\rm F}\sigma}\sim m_{\rm band}/m^{*}\ll 1$. 
Then, the $\bf q$ dependence of 
$\bar{\chi}^{\perp}_{\rm spin}({\bf q},i\omega_0)$ is 
estimated 
as 
$
\bar{\chi}^{\perp}_{\rm spin}({\bf q},i\omega_0)\sim
\bar{\chi}^{\perp}_{\rm spin}({\bf 0},0)\left[
1+\bar{A}
q^2
-i\bar{C}\omega_0/q
\right]
$
with $\bar{A}$ in the order of 
${\cal O}(V_{{\bf k}_{\rm F}}/|\varepsilon_{\rm f}|)^2\lsim 10^{-1}$ 
for typical heavy-electron systems  
with the spherical Fermi surface with $q=|{\bf q}|$~\cite{M07}.  
Then, the most singular term is evaluated 
to show the $(T_1T)^{-1}\sim q^{d-5}$ divergence for $q \to 0$ 
in the $d$-dimensional system. 
This result is in sharp contrast to the single-orbital system 
where charge instability does not occur simultaneously with the spin instability.

The uniform spin susceptibility is defined by 
$
\chi_{\rm s}(T)\equiv
\chi_{\rm s}^{\rm f}(T)+\chi_{\rm s}^{\rm c}(T)
$
with $\chi_{\rm s}^{\rm a}=
\partial m^{\rm a}/\partial H |_{H=0}
$
and $m^{\rm a}\equiv\sum_{i}\langle S^{{\rm a}z}_{i}\rangle/N$, 
when the magnetic field is applied to (\ref{eq:PAM}) as 
$-g_{\rm f}\mu_{\rm B}H\sum_{i}S^{{\rm f}z}_{i}
-g_{\rm c}\mu_{\rm B}H\sum_{i}S^{{\rm c}z}_{i}
$. 
By using the dynamical spin susceptibility 
$
\chi_{+-}^{\rm ab}({\bf q},i\omega_n)
\equiv\int_{0}^{\beta}d\tau
\langle T_{\tau}S^{\rm a}_{+}
({\bf q},\tau)S^{\rm b}_{-}({\bf -q},0)\rangle
{\rm e}^{i\omega_n\tau} 
$, 
$\chi_{\rm s}(T)$ is expressed as 
$
\chi_{\rm s}(T)=\frac{3}{2}\mu_{\rm B}^2
\left[g_{\rm f}^2\chi_{+-}^{\rm ff}
+g_{\rm f}g_{\rm c}\chi_{+-}^{\rm fc}
+g_{\rm c}g_{\rm f}\chi_{+-}^{\rm cf}
+g_{\rm c}^2\chi_{+-}^{\rm cc}
\right], 
$
where ${\bf q}$ and $\omega_n$ are set to be zero 
in the right hand side (r.h.s.).
In heavy-electron systems,
the uniform spin susceptibility is dominated by the f-electron part 
as $\chi^{\rm f}_{\rm s}(0)\gg \chi^{\rm c}_{\rm s}(0)$~\cite{Kyamada}. 
Then, 
$
g_{\rm f}^2\chi_{+-}^{\rm ff}
+g_{\rm f}g_{\rm c}\chi_{+-}^{\rm fc}
\gg
g_{\rm c}g_{\rm f}\chi_{+-}^{\rm cf}
+g_{\rm c}^2\chi_{+-}^{\rm cc}
$
holds. 
Since $\chi_{+-}^{\rm fc}=\chi_{+-}^{\rm cf}$, 
we have
$
\chi_{\rm s}(T)\sim (3/2)\mu_{\rm B}^2g_{\rm f}^2
\chi_{+-}^{\rm ff}
$. 
Hence, at the QCP, $\chi_{\rm s}(0)$ shows 
$\omega_{\rm v}^{-1}$ divergence 
with 
$\omega_{\rm v}\equiv1-\tilde{U}_{\rm fc}
\bar{\chi}^{||}_{\rm spin}({\bf 0},0)$. 

We note here that quite different behavior appears in the charge sector: 
Our DMRG calculation applied to the Hamiltonian (\ref{eq:PAM}) shows 
that even at the QCP 
of the valence transition where 
$\kappa^{\rm f}$ 
diverges, 
the total charge compressibility 
$\kappa_{\rm tot}\equiv\partial \langle n^{\rm f}\rangle/\partial \mu
+\partial \langle n^{\rm c}\rangle/\partial \mu
$ 
does not diverge (see Fig.~\ref{fig:PD}(c)). 
Namely, the r.h.s. of 
$
\kappa_{\rm tot}(0)=
\chi^{\rm ff}
+\chi^{\rm fc}
+\chi^{\rm cf}
+\chi^{\rm cc}
$
are cancelled each other to give a finite value in spite that $\chi^{\rm ff}$ diverges. 
Here, 
${\bf q}$ and $\omega_n$ are set to be zero in 
$
\chi^{\rm ab}({\bf q},i\omega_n)
\equiv\int_{0}^{\beta}d\tau
\langle T_{\tau}n^{\rm a}
({\bf q},\tau)n^{\rm b}({\bf -q},0)\rangle
{\rm e}^{i\omega_n\tau} 
$.  
This is in sharp contrast to the mean-field result, where the first-order valence 
transition is accompanied by the phase separation. 
Namely, diverging relative charge fluctuation (i.e., 
valence fluctuation) also induces the instability of the total charge 
in the mean-field framework. 
However, our finding shows that quantum fluctuations and electron correlations can 
make the total charge stable even though the relative charge unstable.
This is ascribed to the fact that the order parameter of the valence transition, 
$n^{\rm f}=\sum_{i=1}^{N}n^{\rm f}_{i}$ 
is not the conserving quantity, i.e., $[n^{\rm f},H]_{-}\ne 0$~\cite{WIM}. 
This result indicates that the first-order valence transition is not accompanied 
by the phase separation at least in electronic origin.
We point out that the g factors in $\chi_{\rm s}(T)$, 
which usually differs between f and conduction electrons, 
i.e., $g_{\rm f}\ne g_{\rm c}$ prevent from the cancellation 
among $\chi^{\rm ab}_{+-}$ as occurred in the r.h.s. of $\kappa_{\rm tot}(0)$, 
and hence the divergent behavior can emerge in $\chi_{\rm s}(0)$ 
in contrast to $\kappa_{\rm tot}(0)$. 

The enhanced $\chi_{\rm s}(T)$ toward low temperature 
has been observed in $\rm YbAuCu_4$~\cite{Sarrao}. 
The NQR and susceptibility measurements suggest that $\rm YbAuCu_4$ is located near 
the QCP of the valence transition at $H=0$~\cite{wada2}, 
although the antiferromagnetic order masks the ground state with $T_{\rm N}=0.8$~K. 
Namely, the valence-crossover temperature $T_{\rm v}$ is suppressed 
to be close to $T=0$~K so that 
$\chi_{\rm s}(T)$ is interpreted to be enhanced toward low temperature 
by the critical valence fluctuations. 
On the other hand, 
in $\rm YbAgCu_4$ where no magnetic transition has been observed at ambient pressure 
at $H=0$, the lattice constant changes around $T_{\rm v}=40$~K, 
suggesting the valence crossover~\cite{Koyama}. 
A remarkable point is that $\chi_{\rm s}(T)$ has a peak at $T_{\rm v}$~\cite{Sarrao}, 
whose maximum value is one order of magnitude smaller than that of $\rm YbAuCu_4$. 
This is again consistent with our theory, since 
at finite $T_{\rm v}$, i.e., at the valence-crossover point 
$\chi_{\rm s}(T_{\rm v})$ is enhanced but does not diverge,
while in the case of $T_{\rm v}=0$, i.e., at the QCP 
$\chi_{\rm s}(0)$ diverges (see Fig.~\ref{fig:PD}(b) and Fig.~\ref{fig:PD}(d)). 
We note that a recent $(T_1T)^{-1}$ measurement has also detected 
a broad peak around $T_{\rm v}$ in $\rm YbAgCu_4$~\cite{Otani}.  
It is noted that 
the maximum in $\chi_{\rm s}(T)$ has been also observed at $T_{\rm v}$ 
for X=Tl~\cite{Sarrao}. 
Hence, as X moves as Au, Ag to Tl, $T_{\rm v}$ increases, i.e., 
the distance from the QCP becomes long, 
which makes the peak value of the spin susceptibility $\chi_{\rm s}(T_{\rm v})$ 
small. 
Then, our theory explains these systematic observations quite consistently.


Finally, we argue the Wilson ratio near the QCP. 
The self energy for f electrons is given by 
$
\tilde{\Sigma}^{\rm ff}_{\sigma}({\bf p},i\varepsilon_n)=
\frac{T}{N}
\sum_{{\bf q},m}
\frac{2U_{\rm fc}G^{\rm cc}_{\sigma}({\bf p}-{\bf q},i\varepsilon_n-i\omega_{m})}
{1-U_{\rm fc}\bar{\chi}^{\rm fc}_{\sigma\sigma}
({\bf q},i\omega_m)}
$
within the RPA as illustrated in Fig.~\ref{fig:RPA}(f). 
Here, 
$
\bar{\chi}^{\rm fc}_{\sigma\sigma}({\bf q},i\omega_m)
\equiv-\frac{T}{N}\sum_{{\bf k},n}
G^{\rm ff}_{\sigma}({\bf k},i\varepsilon_{n})
G^{\rm cc}_{\sigma}({\bf k}+{\bf q},i\varepsilon_n+i\omega_m) 
$ 
where 
$
G^{\rm ff}_{\sigma}({\bf k},i\varepsilon_n)
=
\bar{a}_{{\bf k}\sigma}^{\rm ff}/
(i\varepsilon_n-E^{*}_{{\bf k}\sigma})+G^{\rm ff}_{\rm inc}
$ 
and 
$G^{\rm cc}_{\sigma}({\bf k},i\varepsilon_n)
=
\bar{a}_{{\bf k}\sigma}^{\rm cc}/
(i\varepsilon_n-E^{*}_{{\bf k}\sigma})+G^{\rm cc}_{\rm inc}$ 
with $U_{\rm fc}$ set to be zero in the previous definition of 
$E^{*}_{{\bf q}\sigma}$. 
Here, 
$
\bar{a}_{{\bf k}\sigma}^{\rm ff}\equiv 
[
1-
\partial \Sigma^{\rm ff}_{{\bf k}\sigma}(\varepsilon)
/\partial\varepsilon|_{\varepsilon=0}
+V_{\bf k}^2
/\varepsilon_{\bf k}^{2}
]^{-1}
$
and 
$
\bar{a}_{{\bf k}\sigma}^{\rm cc}\equiv 
\bar{a}_{{\bf k}\sigma}^{\rm ff}
V_{\bf k}^{2}
/\varepsilon_{\bf k}^{2}
$, 
and 
$G^{\rm ff}_{\rm inc}$ and $G^{\rm cc}_{\rm inc}$ denote the incoherent parts. 
Near the QCP small ${\bf q}$ and $\omega$ components are important 
and the denominator of $\tilde{\Sigma}^{\rm ff}_{\sigma}$ is expanded as 
$1-U_{\rm fc}\bar{\chi}^{\rm fc}_{\sigma\sigma}({\bf q},\omega+i\delta)
\sim
\bar{\omega}_{\rm v}+Aq^2-iC\omega/q$ 
with $\bar{\omega}_{\rm v}\equiv1-U_{\rm fc}\bar{\chi}^{\rm fc}_{\sigma\sigma}({\bf 0},0)$~\cite{M07}. 
Then, 
we evaluate 
$m^*/m_{\rm band}=
1-\partial{\rm Re}\tilde{\Sigma}^{\rm ff}_{\sigma}({\bf p}_{\rm F},\varepsilon)
/\partial\varepsilon|_{\varepsilon=0}=1-
(\bar{a}_{{\bf p}_{\rm F}\sigma}^{\rm cc}U_{\rm fc}/(4\pi^2 v_{\rm F} A))
\ln|\bar{\omega}_{\rm v}
/(\bar{\omega}_{\rm v}+A{q_{\rm c}}^2)|$ to the leading order of 
$\bar{a}_{{\bf p}\sigma}^{\rm cc}\sim \bar{a}_{{{\bf p}_{\rm F}}\sigma}^{\rm cc}$, 
where $v_{\rm F}$ is the Fermi velocity and $q_{\rm c}$ is a cut-off. 
Namely, the Sommerfeld constant $\gamma_{\rm e}$ shows a log divergence 
at the QCP. 
Since $\chi_{\rm s}(0)$ shows a $\bar{\omega}_{\rm v}^{-1}$ divergence, 
the Wilson ratio 
$R_{\rm W}=4\pi^2k_{\rm B}^2\chi_{\rm s}/(3(g_{\rm f}\mu_{\rm B})^2\gamma_{\rm e})$ 
diverges at the QCP of the valence transition $(\bar{\omega}_{\rm v}\to 0)$. 

Although the Gaussian fixed point of the QCP of the valence transition ensures 
the validity of the RPA description~\cite{M07}, 
more qualitatively 
$\bar{\omega}_{\rm v}$ should be determined self-consistently as done 
in the SCR theory for spin fluctuations~\cite{Moriya}. 
This kind of analysis can be executed for valence fluctuations 
starting from the valence susceptibility 
$\chi^{\rm ff}({\bf q},\omega)^{-1}\sim \bar{\omega}_{\rm v}+Aq^2-iC_{q}\omega$, 
expanded near $({\bf q},\omega)=({\bf 0},0)$. 
Here, $C_{q}$ is given, in the $T\to 0$ limit, by the form $C/{\rm max}\{q,l^{-1}\}$ 
with $l$ being the mean free path of the impurity scattering~\cite{MNO}. 
Then, in the realistic situation, the dynamical exponent $z_{\rm d}=3$ is expected to be observed 
at low $T$, except in the very vicinity of $T=0$~K where impurity scattering 
is dominant and hence the system is described by $z_{\rm d}=2$~\cite{note}. 
Since in the three dimension with $z_{\rm d}=3$ the susceptibility and Sommerfeld constant 
behave as $\chi_{\rm s}(T)\sim T^{-4/3}$ and $\gamma_{\rm e}\sim -{\rm ln}T$, 
respectively, 
in case that incoherent part of f electrons gives minor contributions to the criticality, 
the Wilson ratio diverges in this case, too. 
Namely, our results based on the RPA are considered to be qualitatively correct 
and more quantitative arguments for comparison with experiments 
will be reported in the separated paper~\cite{WM2}. 

We note that 
$\chi_{\rm s}(0)$ and $\gamma_{\rm e}$ extrapolated 
to $T\to 0$~K from the $T>T_{\rm N}=0.8$~K data in $\rm YbAuCu_4$ 
shows $R_{\rm W}\sim 4.5$~\cite{Sarrao}, 
which exceeds $R_{\rm W}=2$. 
Enhanced Wilson ratio $R_{\rm W}\sim 3$ 
has been also observed 
in ${\rm Ce}_{0.9-x}{\rm La}_{x}{\rm Th}_{0.1}$ with $x=0.1$~\cite{Lashley} 
where the $\gamma$-$\alpha$-transition temperature is suppressed 
closely to zero temperature. 
These materials are located near the QCP of the valence transition 
and hence our theory gives an explanation for these enhancements. 
Enhanced $R_{\rm W}$ has been also observed in other paramagnetic materials such as 
$\rm YbRh_2(Si_{0.95}Ge_{0.05})_2$~\cite{YbRh2Si2}, $\rm YbIr_2Si_2$~\cite{YbIr2Si2} 
and $\beta$-$\rm YbAlB_4$~\cite{nakatuji}, 
which also suggests underlying influence of valence fluctuations. 

We stress that our results can be generally applied to the systems 
with valence instabilities. Experimental examination of our predictions is highly desired.

The authors thank S. Wada and S. Nakatsuji for 
showing us their experimental data prior to publication.


\begin{thebibliography}{99}
\bibitem{chipm} Even when $\bar{\chi}_{\rm charge}$ is not negligible for 
$\bar{\chi}_{\rm spin}^{||}$, $\chi^{\rm ff}_{+-}({\bf 0},0)$ is expected to be greatly enhanced at 
$T_{\rm v}$ as well as the critical points.
\end{thebibliography}

\end{document}